\documentclass{SCGE}

\begin{document}

\begin{picture}(0,0){\rm
\put(0,-20){\makebox[160truemm][l]{\bf {\sanhao\raisebox{2pt}{.}}
Article  {\sanhao\raisebox{1.5pt}{.}}}}}
\put(0,-34){\jiuwuhao {\textcolor[rgb]{0.5,0.5,0.5}{\sf 
}}}
\end{picture}

\def\bm{\boldsymbol}

\def\dl{\displaystyle}
\def\du{\end{document}}
\def\d{{\rm d}}
\def\e{{\rm e}}
\def\i{{\rm i}}
\def\pi{{\uppi}}

\Year{2015} %
\Month{10} %
\Vol{58} 
\No{10} 
\BeginPage{1} 
\AuthorMark{{\rm Liu B}, et al.}  
\AuthorMarkCite{{\rm Liu B, Gao F, Huang W}, et al. } 
\DOI{10.1007/s11433-015-5714-3} 
\ArtNo{100301}

\title[QKD-based quantum private query without a failure probability]
{QKD-based quantum private query without a failure probability}

\author[1,2]{LIU Bin}{}
\author[1,*]{GAO Fei}{}
\author[1]{HUANG Wei}{}
\author[1]{WEN QiaoYan}{}

\address[{\rm1}]{State Key Laboratory of Networking and Switching Technology, Beijing University of Posts and Telecommunications, Beijing, 100876, China}
\address[{\rm2}]{College of Computer Science, Chongqing University, Chongqing 400044, China}

\maketitle \vspace{-3.5mm}{\footnotesize\begin{center} Received June 2, 2015; accepted June 18, 2015
\end{center}}\vspace*{-5mm}

\begin{center}
\rule{16.5cm}{0.4pt}
\parbox{16.5cm}
{\begin{abstract}
In this paper, we present a quantum-key-distribution (QKD)-based quantum private query (QPQ) protocol utilizing single-photon signal of multiple optical pulses. It maintains the advantages of the QKD-based QPQ, i.e., easy to implement and loss tolerant. In addition, different from the situations in the previous QKD-based QPQ protocols, in our protocol, the number of the items an honest user will obtain is always one and the failure probability is always zero. This characteristic not only improves the stability (in the sense that, ignoring the noise and the attack, the protocol would always succeed), but also benefits the privacy of the database (since the database will no more reveal additional secrets to the honest users). Furthermore, for the user's privacy, the proposed protocol is cheat sensitive, and for security of the database, we obtain an upper bound for the leaked information of the database in theory.
\end{abstract}}
\end{center}\vspace*{-0.6cm}

\begin{center}
\parbox{16.5cm}
{\bf\jiuhao quantum private query, quantum key distribution, single-photon signal}
\end{center}

\begin{center}
{}
\CITA    
\end{center}

\textwidth=178truemm \textheight=236truemm

\wuhao\vspace*{1.5mm}

\begin{multicols}{2}

\renewcommand{\baselinestretch}{1.08} \baselineskip 12.2pt\parindent=10.8pt

\renewcommand{\thefootnote}

\section{Introduction}

The applications of quantum mechanics have achieved great success in mutual trusted private communication, for which quantum key distribution (QKD) can allow two parties who trust each other to generate a secret key over a public channel with ``unconditional'' security \cite{BB84,Lo1999,Shor2000}. In the last three decades, mutual trusted private communication has been fully developed both in theory\cite{BB84,Lo1999,Shor2000,E91,B92,GV95,T6,T7,T8,T9,T10,T11} and in experiment\cite{E5,E6}. While in some cryptographic communications, we need not only protect the security of the transmitted message against the outside adversaries, but also the communicators' individual privacies against each other \cite{LC1997,may1997,G1,H2}. An interesting example is the problem of private query to a database, where Alice (the user) wants to get an item in Bob's database, which is composed of a quantity of secret messages. In this scenario, Alice does not want Bob to know which item she is interested in and Bob does not want Alice to get more items of his database.
Symmetrically private information retrieval (SPIR) \cite{SPIR} protocols are designed for such circumstance. It ensures the privacies of both Alice and Bob. That is, after the communication, Alice can correctly obtain precisely one item (i.e., the one Alice wants to get) from Bob's database and simultaneously, Bob does not know the address Alice has retrieved.

Quantum private query (QPQ) is the quantum scheme for SPIR problem. Similar to the two-party quantum secure computations, the task of SPIR cannot be achieved ideally even in its quantum version \cite{Lo1997}. More practically, the aim of quantum private query is generally loosened into that Alice can elicit a few more items than the ideal requirement (i.e, just one item) in the database, and Bob's attack will be discovered with a non-zero probability by Alice if he tries to obtain the address Alice is retrieving (equivalently, Alice's privacy is guaranteed in the sense of cheat-sensitive instead of an ideal one).

In 2008 V. Giovannetti et al. proposed the first quantum private query protocol \cite{QPQ08,QPQproof,QPQexp}, where the database is represented by an unitary operation (i.e. oracle operation) and it is performed on the two coming query states. In 2011 L. Olejnik presented an improved protocol where only one query state is needed \cite{QPIR11}. Compared with the previous (classical) SPIR schemes, the above two protocols not only exhibit advantage in security (i.e. security based on fundamental physical principles instead of assumptions on computational complexity) but also display an exponential reduction in both communication complexity and running-time computational complexity.

Though the above two protocols have significant advantages in theory, they are difficult to implement because when large database is concerned, the dimension of the oracle operation will be very high. To solve this problem, M. Jakobi et al. gave a new private query protocol based on quantum oblivious key transfer (QOKT) \cite{QPQ11}.
This type of QPQ protocol is called QKD-based QPQ \cite{QPQ11} or QOKT-based QPQ \cite{GaoLiu15}. The previous QKD-based QPQ protocols can be generally divided into the following three Stages.

(1) \textbf{States transfer} (Raw oblivious key distribution). Alice and Bob generate a raw oblivious key satisfying: (a) Bob knows the whole key; (b) Alice only knows part of it; (c) Bob does not know which bits Alice knows.

(2) \textbf{Post-processing} (Oblivious key dilution). Alice and Bob ``dilute'' the raw oblivious key into a final oblivious key satisfying: (a) Bob knows the whole key; (b) Alice knows only one (or a little more than one) bit; (c) Bob does not know which bit Alice knows. \textit{Note that the key dilution stage in all the previous QKD-based QPQ protocols would fail with a small probability and if it happens, Alice and Bob have to restart the whole protocol. And generally, to make the above failure probability smaller, the expectation of the number of the bits Alice gains is always larger than one, which is actually not in Bob's favor.}

(3) \textbf{Private Query}. Alice claims a shift to the final key according to the item she wants to query. Then Bob encrypts his database by the shifted final oblivious key and sends the whole encrypted database to Alice. Thus Alice will know exactly the bit she is interested in.

Compared with the previous QPQ protocols, the above type of QPQ protocol, i.e., QKD-based QPQ, is easy to be realized. More concretely, it can be easily generalized to large database, and it is loss tolerant.
Therefore, as a practical model, QKD-based QPQ is very attractive and has become a research hotspot. Different manners to distribute the raw key were given in Refs. \cite{OE12,PRA13,Wei14,YYG14,YYG142,YYG15,YYG152}, and different methods for post-processing were presented in Refs. \cite{PJ13,SZM12} and analyzed by us in Ref. \cite{GaoLiu15}. What's more, recently, people start to study the error-correction scheme for the QOKT \cite{GaoLiu15,otherEC}.

Inspired by a recently proposed QKD protocol \cite{nature14} utilizing single-photon signal of multiple optical pulses, we propose a QPQ protocol here. As a QKD-based QPQ protocol, it maintains the original advantages, i.e., easy to implement and loss tolerant. (Note that the single-photon signal of multiple optical pulses sources have good experimental foundation, for instance the multi-mode W-state has been generated and precisely characterised \cite{C1,C2}). Besides, different from the situation in the previous QKD-based QPQ protocols, the number of the bits an honest user (Alice) knows in the initially generated oblivious key (i.e., the raw oblivious key in the previous protocols) is always one. Therefore, no more key dilution process is required here, and consequently there is no more failure probability for our protocol. This characteristic not only improves the stability, in the sense that, ignoring the noise and the attack, the protocol would always succeed, but also benefits the privacy of the database since the database owner (Bob) will no more reveal additional secrets to the honest users (Alice).
Moreover, in all the previous QKD-based QPQ protocols, only some specific attacks to the database are analyzed. Differently, here we calculate an upper bound for the leaked information of the database in the proposed protocol in theory.
What's more, like all the previous QPQ protocols (not only the QKD-based ones), our protocol is also cheat sensitive, in the sense that if Bob tries to steal the information on which item Alice is querying, he will not give Alice the item precisely and, therefore, Alice could detect Bob's cheating by the incorrect data value.

The rest of this paper is organized as follows. The protocol is presented in next section. In section 3, we analyze both the database security and the user security of the proposed protocol. And a short conclusion is given in section 4.

\section{Quantum Private Query Utilizing Single-Photon Signal}
\subsection{Bit sharing process utilizing single-photon signal of multiple pulses}
Recently, utilizing a single-photon state of $L$ optical pulses, Sasaki et al. proposed a QKD protocol without monitoring signal disturbance, which has high tolerance of bit errors. Here we introduce a variant of the bit sharing process in Sasaki et al. protocol, where the realization condition is a little more idealized than that of the original version.
\begin{itemize}
  \item [a] Bob generates a $L$-length binary string $S$ and a single-photon state of $L$ optical pulses $|\Psi_0\rangle$, where the only photon might be equiprobably in each of the pulses. Here,
      \begin{eqnarray}\label{S}
        S=s_0s_1\cdots s_{L-1},
        \end{eqnarray}
        where $s_i=0$ or $1, i=0,1,\cdots,L$$-$$1$; and
        \begin{eqnarray}\label{S}
        |\Psi_0\rangle=\frac{1}{\sqrt{L}}\sum\limits_{{k} = 0}^{L-1}{|k\rangle},
      \end{eqnarray}
  where the photon is in the $k$-th pulse for state $|k-1\rangle$.
  Then according to $S$, Bob modulates the phase of each pulse and the state changes into
  \begin{equation}\label{psis}
    |\Psi_S\rangle=\frac{1}{\sqrt{L}}\sum\limits_{{k} = 0}^{L-1}{(-1)^{s_k}|k\rangle}.
  \end{equation}
  Then Bob sends $|\Psi_S\rangle$ to Alice.
  \item [b] Alice generates a random value $r\in \{1,2,\cdots L$$-$$1\}$ in advance. When she receives $|\Psi_S\rangle$, she first splits each pulse by a half beam-splitter and then, according to $r$, she changes the order of the pulses in one of the light paths into
      \begin{equation}
      r, r+1, \cdots, L-1, 0, 1, \cdots, r-1,
      \end{equation}
      while the order of the pulses in the other light path remains
      \begin{equation}
      0, 1, \cdots, r-1, r, r+1, \cdots, L-1.
      \end{equation}
      Then she superposes the pulses in the two light paths in order. According to the measurement result (i.e., in which pair of pulses he denotes an photon and the interference result), she will get one of the following values (if no eavesdropping and noise exist)
      \begin{equation}
        s_{i\oplus_Lr}\oplus s_i,
      \end{equation}
      where $i=0,1,\cdots L$$-$$1$, $\oplus$ denotes summation modulo 2 and $\oplus_L$ denotes summation modulo $L$.
  \item [c] If Alice gets the interference result of the $(i_0$+1)-th pair of the pulses (i.e., the value of  $s_{i_0\oplus_Lr}\oplus s_{i_0}$,), she publishes $i_0$ and $r$. Alice and Bob agree on that the shared key bit in this round is just $s_{i_0\oplus_Lr}\oplus s_{i_0}$.
\end{itemize}

We denote the processes presented above as Protocol I. The more concrete content about Sasaki et al. QKD protocol, for example the security analysis and the implementation scheme in practice, please refer to the Ref. \cite{nature14}. Here we will not dwell on it since this paper only covers this novel coding scheme described above.

\subsection{The proposed oblivious key distribution protocol}
Utilizing single-photon signal of multiple pulses, we can distribute a oblivious key which is very suitable for the QPQ problem.
Our QPQ protocol (denoted as Protocol II) for an $N$-element database utilizing single-photon signal of multiple optical pulses is as follows.

\begin{itemize}
  \item [1] Bob generates a single-photon state of $N$+1 optical pulses $|\Psi_{S}\rangle$ according to an ($N$+1)-bit binary string $S$ (just like the state in Eq. \ref{psis}),
  \begin{equation}\label{eq22}
    S=s_{0},s_{1},s_{2},\cdots,s_{N},
  \end{equation}
and
  \begin{eqnarray}\label{phiSk}
        |\Psi_{S}\rangle=\frac{1}{\sqrt{N+1}}\sum\limits_{{k} = 0}^{N}{(-1)^{s_{k}}|k\rangle},
  \end{eqnarray}
  where $s_{k}$ is randomly 0 or 1, and $k$=$0,1,2,\cdots N$. Then he sends  $|\Psi_{S}\rangle$ to Alice in sequence. 
  \item [2] Alice generates a random value $r\in \{1,2,\cdots N\}$, and then she uses the interference circuits in Fig. \ref{liucheng} (just as what Alice does in Step b in Protocol I) to randomly get one of the following values
      \begin{equation}
        s_{j} \oplus s_{j\boxplus r},
      \end{equation}
      where $j=0,1,\cdots N$, $\oplus$ denotes summation modulo 2 and $\boxplus$ denotes summation modulo $N$+1. For ease of description, we suppose the value Alice finally gets is
      \begin{equation}\label{a}
        s_{t}\oplus s_{t\boxplus r}.
      \end{equation}
       Note that Alice got the interference result of the ($t$+1)-th pair of the pulses in two paths.
\begin{figure}[H]
  \includegraphics[scale=0.5]{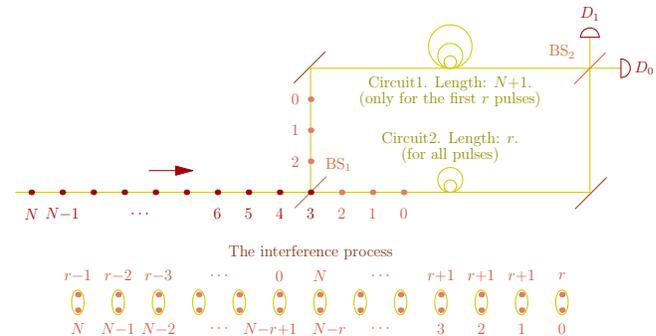}\\
  \caption{Alice's information extraction process: After splitting the pulses of $|\Psi_{S}\rangle$, Alice sets up two circuits in the two light paths according to $r$, respectively. One, Circuit 1, is with the length of $N+1$ pulses and this circuit only works for the first $r$ pulses, i.e., the rest pulses will take shortcut without Circuit 1. The other, Circuit 2, is with the length of $r$ pulses and all pulses in this light path will go through it. Thus, Alice can get interference results of the two-way pulses in the way shown in the bottom half of the figure.}\label{liucheng}
\end{figure}
  \item[3] Alice publishes $t$ and they agree on the $N$-bit key
  \begin{eqnarray}\label{ObK}
    s_{t}\oplus s_{0}, s_{t}\oplus s_{1}, \cdots, s_{t}\oplus s_{t-1},\nonumber\\
     s_{t}\oplus s_{t+1}, \cdots, s_{t}\oplus s_{N-1}, s_{t}\oplus s_{N}.
  \end{eqnarray}
  Note that Alice knows $s_{t}\oplus s_{t\boxplus r}$.
  \item [4] According to the position of the bit Alice knows in the $N$-bit key above and the position of the item she wants to query, Alice claims a shift to the key. For example, if Alice knows the $j$-th bit in the key and wants to retrieve the $i$-th item in Bob's database, she declares a shift value $s$$=$$j$$-$$i$ so that Bob can shift his key by $s$.
  \item [5] Bob shifts the key according to the value Alice declared. Then he encrypts his database by the shifted oblivious key and sends the encrypted database to Alice.
  \item [6] Alice decodes the item she wants to query by the shifted oblivious key.
\end{itemize}

In Protocol II, Alice always knows precisely one bit in the oblivious key, therefore, the protocol would always succeed if no attack and noise exist, and Bob no more needs to leak additional information about his database to the honest Alice.

In addition, to verify Bob's honesty, Alice could periodically check the validity of the received data in other database, or query to the same item repeatedly and compare the outcomes \cite{QPIR11}.
Another way, similar to the first QPQ protocol\cite{QPQ08}, they would perform Protocol II twice at one query process, denoted as Protocol II$^\prime$, where Alice is allowed to query the same item twice to check Bob's honesty.
In fact, the two ways to verify Bob's honesty above are both based on that if Bob gains information on the bit position Alice knows in the key, he will lose information on the bit value she has recorded \cite{QPQ11}. And the latter (Protocol II$^\prime$) appears more active since Alice might detect Bob's dishonest action immediately, however, it leaks more information on the database to the dishonest Alice.

\subsection{Correctness of the proposed protocol}

Now we analyze how each single-photon signal of multiple optical pulses Bob sends to Alice evolves in the whole processes of Protocol II. The initial state is as that in Eq. \ref{phiSk}.
When the signal passed Alice's first half beam-splitter BS$_1$, it changed into
\begin{equation}\label{HB}
    \frac{1}{{\sqrt {2(N + 1)} }}\left[ {\sum\limits_{k = 0}^N {{{( - 1)}^{{s_{k}}}}\left| k \right\rangle } \left| 0 \right\rangle  + \text{i}\sum\limits_{k = 0}^N {{{( - 1)}^{{s_{k}}}}\left| k \right\rangle } \left| 1 \right\rangle } \right],
\end{equation}
where the state $|0\rangle$ represents the photon is in the transmission path and $|1\rangle$ represents the photon is in the reflection path. After passing the two delay circuits and the two reflectors, it becomes
\begin{equation}\label{DC}
    \frac{\text{i}}{{\sqrt {2(N + 1)} }}\left[ {\sum\limits_{k = 0}^N {{{( - 1)}^{{s_{k}}}}\left| k \right\rangle } \left| 0 \right\rangle  + \text{i}\sum\limits_{k = 0}^N {{{( - 1)}^{{s_{k \boxplus {r}}}}}\left| k \right\rangle } \left| 1 \right\rangle } \right].
\end{equation}
After the two shifted pulses passes through the second half beam-splitter BS$_2$, the state becomes
\begin{equation}\label{MS}
 \frac{\text{i}}{{2\sqrt {(N + 1)} }}\left[ \begin{array}{l}
\sum\limits_{k = 0}^N {\left[ {{{( - 1)}^{{s_{k}}}} - {{( - 1)}^{{s_{i,k \boxplus {r}}}}}} \right]\left| k \right\rangle } \left| 1^\prime \right\rangle \\
 + \text{i}\sum\limits_{k = 0}^N {\left[ {{{( - 1)}^{{s_{k}}}} + {{( - 1)}^{{s_{k \boxplus {r}}}}}} \right]\left| k \right\rangle } \left| 0^\prime \right\rangle
\end{array} \right],
\end{equation}
where the state $|0^\prime\rangle$ represents the photon is in the light path leading to the detector $D_0$ (i.e., the transmission-reflection path and also the reflection-transmission path), and $|1^\prime\rangle$ represents the photon is in the light path leading to the detector $D_0$ (the transmission-transmission path and also the reflection-reflection path).
We can see that if $s_{k}$$=$$s_{k\boxplus r}$, Alice might detect the photon in $D_0$ but never detect it in $D_1$, and correspondingly if $s_{k}$$\neq$$s_{k\boxplus r}$, she might detect the photon in $D_1$ but never detect it in $D_0$. Therefore, once Alice detect the photon at $(t$+1)-th pair of the pulses, she can know the value of $s_{t}\oplus s_{t\boxplus r}$.

\section{Security analysis of the proposed protocol}

\subsection{Database security}
It is difficult to figure out a specific attack strategy for Alice to steal more information of Bob's database. Here we calculate an upper bound of a dishonest Alice's information about Bob's database by the Holevo bound. That is
\begin{align}\label{holevo}
    &\ \ \ H(A:\text{Database})\leq H(A:S)\\
    &\leq S(\frac{1}{{{2^{N + 1}}}}\sum\limits_{k = 1}^{{2^{N + 1}}} {\left|\Psi_k \right\rangle \left\langle \Psi_k \right|} ) - \frac{1}{{{2^{N + 1}}}}\sum\limits_{k = 1}^{{2^{N + 1}}} {S(\left| \Psi_k \right\rangle \left\langle \Psi_k \right|)}\nonumber
\end{align}

Here, $S(\rho)$ represents the Von Neumann entropy of the state $\rho$, $H(A:\text{Database})$ denotes the amount of information that Alice can gain about Bob's database, $H(A:S)$ denotes the amount of information that Alice can gain about Bob's binary strings $S$, and each $|\Psi_k \rangle$ denotes one of the $2^{N+1}$ different scenarios of the states in Eq. \ref{phiSk}. And we have
\begin{eqnarray}
 S(\frac{1}{2^{N + 1}}\sum\limits_{k = 1}^{2^{N + 1}} {\left|\Psi_k \right\rangle \left\langle \Psi_k \right|})=S(I_{N+1})=\log(N+1),
 \end{eqnarray}
 \begin{eqnarray}
  S(\left| \Psi_k \right\rangle \left\langle \Psi_k \right|)=0.
\end{eqnarray}
Therefore,
\begin{equation}\label{holevo1}
    H(A:\text{Database})\leq H(A:S)\leq \log(N+1).
\end{equation}
(Obviously, $H(A:\text{Database})\leq 2\log(N+1)$ for Protocol II$^\prime$.) This upper bound is lower than that of the previous QKD-based QPQ protocols when a very large database is considered.
For example in the QPQ protocol in Refs. \cite{QPQ11,OE12}, the average information Alice can get for each bit in the final key can be described by a constant $\delta$ if Alice stores the qubits corresponding to the same bit in the final key and performs collective measurements to them (which is just a specific attack strategy and there might be some more powerfull attacks, for example, to perform collective measurements to the whole received photons together), and consequently she can get $N\delta$ information about Bob's database in total. And when $N$ become larger, the advantage of our protocol's bound $B_1\leq \log(N+1)$ would be highlighted comparing with $B_2 \geq N\delta$ in Ref.s \cite{QPQ11,OE12}.

\subsection{User privacy}

Just as in the previous QPQ protocols \cite{QPQ08,QPQ11,QPIR11,OE12}, the security of user's privacy in our protocols is based on that if Bob gets some effective information on which item Alice queries, he might give Alice the wrong value of the queried one, and Alice would detect Bob's attack by the incorrect item.

Firstly, we consider that Bob sends Alice a pure state of single photon with $N+1$ pulses as follows
\begin{equation}\label{AS}
|\Psi_A\rangle=\sum\limits_{k = 0}^N {{a_k}\left| k \right\rangle } ,
\end{equation}
where $\sum_k {|{a_k}{|^2}}  = 1$. (The signals with 2 or more photons will be detect by Alice's interference circuits.)
When the above signal passes the two half beam-splitters, it becomes
\begin{equation}\label{AS2}
    \frac{1}{2}\sum\limits_{k = 0}^N {\left[ {({a_k} + {a_{k \boxplus r}})\left| k \right\rangle \left| 0^\prime \right\rangle  + \text{i}({a_k} - {a_{k \boxplus r}})\left| k \right\rangle \left| 1^\prime \right\rangle } \right]}.
\end{equation}
Thus, if Alice chooses $r$, she will get the photon at the ($k$+1)-th pair of pulses with the probability
\begin{equation}\label{Pk}
   \frac{1}{2}(|{a_k}{|^2} + |{a_{k \boxplus r}}{|^2}).
\end{equation}
The joint probability distribution of different $r$ and $k$ is
\begin{equation}\label{jpd}
   \left[\frac{1}{2N}(|{a_k}{|^2} + |{a_{k \boxplus r}}{|^2})\right]_{r,k},
\end{equation}
where $r$=$1,2,\cdots,N$ and $k$=$0,1,2,\cdots,N$. And the minimum probability of Bob giving a wrong value to Alice is
\begin{eqnarray}\label{mp}
P_{\text{min}}=\sum\limits_{k = 0}^N  {\min \left\{ \frac{{|{a_k} + {a_{k \boxplus r}}{|^2}}}{{4N}},\frac{{|{a_k} - {a_{k \boxplus r}}{|^2}}}{{4N}}\right\}}.
\end{eqnarray}
Obviously, $|\Psi_A\rangle$ should be prepared with the condition $a_i=\pm a_j$ for all $i$ and $j$, if Bob wants to give Alice the item she is querying precisely. Ignoring the overall phase, it is just the legal state $|\Psi_S\rangle$ in Eq. \ref{phiSk}.

Now we consider that Bob prepares an entangled state and sends part of it to Alice. To give Alice the correct item, Bob must know every phase difference $s_t\oplus s_j$ where $j=0,1, \ldots, N$, since $r$ is generated randomly by Alice and $j$ could be any of the natural numbers no bigger than $N$ except $t$. Therefore, the state he sends to Alice must in the following form
\begin{equation}\label{df}
    \sum\limits_{i=1}^{2^{N}}\lambda_i|\Lambda_i\rangle_B|\Psi_{S_i}\rangle_A,
\end{equation}
where $\sum_i |\lambda_i|^2=1$,
\begin{equation}\label{sf}
    |\Psi_{S_i}\rangle=\frac{1}{\sqrt{N+1}}\left(|0\rangle+\sum\limits_{k=1}^{N}{(-1)^{s_{i,k}}|k\rangle}\right),
\end{equation}
$\langle \Psi_{S_i}|\Psi_{S_j}\rangle\neq1$ for $i\neq j$, and $\langle\Lambda_i|\Lambda_j\rangle=\sigma_{ij}$. In the above equation, each value of $i$ is corresponding to one of the $2^N$ different states $|\Psi_{S_i}\rangle$. Note that, ignoring the overall phase, $|\Psi_{S_i}\rangle$ is the same with
\begin{equation}\label{sf1}
    |\Psi^\prime_{S_i}\rangle=\frac{1}{\sqrt{N+1}}\left(-|0\rangle+\sum\limits_{k=1}^{N}{(-1)^{s_{i,k}\oplus1}|k\rangle}\right).
\end{equation}
Then Bob sends Alice the second half of the state in Eq. \ref{df}, i.e., system $A$, and keeps the first, i.e., system $B$. After Alice has measured $A$, he can measure $B$ in bases $\{|\Lambda_i\rangle\}_{i=1}^N$ to get the value of $S_i$, according to which he can give the correct value of the item she wants to query. However, by doing this, he will get no information about which item Alice has queried. Or he can make other measurements on system $B$ to steal the above position information, but thus he cannot get the whole information about $S_i$ and, therefore, he cannot give Alice the correct value all the time. So our protocol is also cheat sensitive just as the previous ones.

For example, Bob may prepare the entangled state
\begin{equation}\label{infse}
    \frac{\sum\limits_{s_0,s_1,\ldots,s_N}|s_0\rangle_0|s_1\rangle_1\ldots|s_N\rangle_N\left(\sum_{k=0}^{N}(-1)^{s_k}|k\rangle_A\right)}{\sqrt{(N+1)\times2^N}},
\end{equation}
where Bob keeps the first $N$$+$$1$ systems $\{0,1,\ldots, N\}$ and sends system $A$ to Alice.
When Alice has measured system $A$ and published part of the measurement results $t$ in Step 2 and 3, the whole system for Bob collapsed into a mixture state with the ensembles of pure states
\begin{eqnarray}\label{xz}
    \{\frac{1}{2},\ \  |\varphi^+\rangle_{t,t\boxplus r}\bigotimes\limits_{\begin{array}{l}_{\ i\neq t,}\\ _{i\neq t\boxplus r}\end{array}}|+\rangle_i\otimes|t0^\prime\rangle_A;\nonumber\\
    \frac{1}{2},\ \  |\psi^+\rangle_{t,t\boxplus r}\bigotimes\limits_{\begin{array}{l}_{\ i\neq t,}\\ _{i\neq t\boxplus r}\end{array}}|+\rangle_i\otimes|t1^\prime\rangle_A\},
\end{eqnarray}
where,
\begin{eqnarray}
  |+\rangle=\frac{1}{\sqrt{2}}\left(|0\rangle+|1\rangle\right),\\
  |\varphi^+\rangle=\frac{1}{\sqrt{2}}\left(|00\rangle+|11\rangle\right),\\
  |\psi^+\rangle=\frac{1}{\sqrt{2}}\left(|01\rangle+|10\rangle\right).
\end{eqnarray}
If Bob measures the first $N$$+$1 system in $Z$ basis, he can give the correct item to Alice. While he can measure them in $X$ basis, and if he gets the result $|-\rangle$$=$$1/\sqrt{2}(|0\rangle-|1\rangle)$ in system $t\boxplus r$, with the probability $1/2$, he would get Alice's privacy. Thus, he will totally lose the information on the bit value of the key. And Alice would detect Bob's cheating with the probability 1/2. 

\section{Summary and conclusion}

In this paper, utilizing the new encoding style by single photon for multiple optical pulses, we proposed a QKD-based QPQ protocol, which, we think, has the following advantages.

\begin{itemize}
  \item It maintains the advantages of the QKD-based QPQ, i.e., easy to implement and loss tolerant.
  \item The failure probability is always 0. The proposed protocol is more reliable compared with the previous QKD-based QPQ protocols, in the sense that, ignoring the noise and the attack, it would always succeed.
  \item The number of items an honest user will obtain is always 1. Therefore, it is more reasonable compared with the previous QKD-based QPQ protocols, since Bob will no more reveal unexpected secrets to the honest Alice.
\end{itemize}

For the security of the database, we have calculated an upper bound for the leaked information of the database in theory. This upper bound is lower than the previous QKD-based QPQ protocols, however, it is larger than that of the ones in Refs. \cite{QPQ08,QPIR11}. And correspondingly, the implementation difficulty of our protocols is also between them. That is, Bob does not need to perform a collective unitary operation on a very large quantum system, however, the single photon signal with $N$$+$1 pulses is more difficult to prepare than the single photons.
For the privacy of the user, our protocols are also cheat sensitive as the previous ones, since once Bob tries to get the address of the item which Alice is querying, he cannot give the correct item to Alice precisely and Alice could detect Bob's cheating by the incorrect data value.

In conclusion, the proposed protocol is the first QKD-based QPQ protocol without the process of the oblivious key dilution, and, therefore, it is the first QKD-based one with no failure probability and no information reveal for the database when the user is honest. Therefore, we think the proposed protocol initiates a new branch of QKD-based QPQ.
And we hope that the contents of this paper would be both interesting and instructive to the researchers who are working on quantum private query and the related fields.

\vspace*{2mm} \Acknowledgements{\bahao This work is supported by NSFC (Grant Nos. 61272057, 61170270), Beijing Higher Education Young Elite Teacher Project (Grant Nos. YETP0475, YETP0477), BUPT Excellent Ph.D. Students Foundation (Grant CX201442).}

\end{multicols}

\end{document}